\documentclass[a4paper,11pt]{article}
\pdfoutput=1 % if your are submitting a pdflatex (i.e. if you have
\usepackage{jcappub} % for details on the use of the package, please
\usepackage{multicol}
\usepackage{multirow}
\usepackage[T1]{fontenc} % if needed
\usepackage{graphicx}
\usepackage[export]{adjustbox}
\usepackage{amsmath}
\usepackage{array}
\usepackage{subcaption}
\usepackage{amsfonts}
\usepackage{amssymb}
\usepackage[left=2cm,right=2cm,top=2cm,bottom=2cm]{geometry}
\usepackage{array}
\usepackage{multirow}
\usepackage{multicol}
\usepackage{tablefootnote}
\usepackage{hyperref}
\usepackage[normalem]{ulem} % Load ulem package

\usepackage{xcolor} % Package to handle colors
\definecolor{DeepGreen}{RGB}{0,150,150} % Dark green

\title {\boldmath From metallicity distributions to mutual information: A new perspective on stellar halo assembly}

\author[a]{Amit Mondal}

\author[a]{and Biswajit Pandey}

\affiliation[a]{Department of Physics, Visva-Bharati University, Santiniketan, 731235, India}

\emailAdd{amitmondal.bwn95@gmail.com} \emailAdd{biswap@visva-bharati.ac.in} 

\abstract{The metallicity structure of stellar halos encodes the fossil record of galaxy assembly, tracing the chemical evolution and dynamical imprint of past mergers. Using five Milky Way-mass halos from the Aquarius simulations, we introduce an information-theoretic framework to quantify spatial-chemical correlations through the mutual information (MI) between angular position and metallicity. We divide stars in each halo into high- and low-metallicity populations based on their median metallicity and examine their metallicity distribution functions (MDFs), spatial anisotropies, and angular-metallicity couplings as a function of galactocentric radius. The MDFs exhibit remarkable diversity, ranging from single-peaked distributions dominated by one or two massive progenitors to broad or bimodal forms shaped by multiple accretion events, revealing the stochastic nature of halo assembly. The low-metallicity stars, primarily contributed by disrupted satellites, display higher spatial anisotropy and stronger angular clustering than their metal-rich counterparts. After removing bound satellites, anisotropy decreases significantly, yet high-metallicity stars remain marginally more anisotropic, reflecting the lingering debris of massive, centrally deposited progenitors. The mutual information between angular position and metallicity increases with radius before saturating in the outskirts, with the difference between the data and randomized controls confined mainly to the inner halo signifying residual spatial-chemical coupling driven by incomplete phase mixing. Our results demonstrate that information-theoretic diagnostics provide a powerful and intuitive way to quantify the chemical complexity of stellar halos and offer a promising route to compare simulations with forthcoming high-dimensional Galactic survey data.}

\begin{document}
\maketitle
\flushbottom

%\begin{keywords}
%methods: statistical - data analysis - galaxies: formation - evolution
%- cosmology: large scale structure of the Universe.
%\end{keywords}

\section{Introduction}

Understanding how metals are distributed within galaxies and particularly within their stellar halos provides vital clues to the processes of galaxy assembly, feedback, and accretion. Spatial variations in metallicity carry the imprint of hierarchical building blocks, gas flows, and internal mixing, and hence offer a fossil record of formation pathways \citep{lian23, monachesi15, dsouza18, gonza25}. The metallicity gradient is a crucial diagnostic which describes the change in the average metallicity of a stellar population as a function of its distance from the galactic center. The presence, absence, or steepness of this gradient provides critical insight into how the stellar populations are spatially distributed and how the chemical enrichment process unfolded within the galaxy's evolving potential well. A steep negative gradient might suggest that the metal-poor stars are preferentially located in the outer halo whereas a flat gradient could indicate a well-mixed or homogenized population.

A wealth of recent theoretical and observational work has established that the metallicity structure of stellar halos is closely tied to their formation history. High-resolution cosmological hydrodynamical simulations such as Illustris \citep{dsouza18}, Auriga \citep{monachesi19}, ARTEMIS \citep{font20}, and CIELO \citep{gonza25} consistently show that halo metallicity is linked to the mass and timing of its dominant accretion events. In Illustris, \cite{dsouza18} demonstrated a tight correlation between stellar halo metallicity and the mass of the most massive accreted progenitor. The scatter in this relation encodes the progenitor’s stellar mass, while the steepness of radial metallicity gradients reflects its accretion time. These trends imply that metallicity is not merely a passive tracer of enrichment but a direct diagnostic of the galaxy’s merger history. 

Simulations also highlight the complexity of halo metallicity profiles. Auriga \citep{grand17} and ARTEMIS \citep{font20} find that spherically averaged metallicity gradients tend to be mild, but can become nearly flat along the minor axis \citep{font20, monachesi15}, particularly in halos with significant in-situ components. Observations paint a similarly diverse picture: some Milky Way-mass galaxies exhibit strong negative gradients \citep{xue15, dietz20} while others show weak or none \citep{dejong10, fernandez15, monachesi16}. Large surveys such as SEGUE \citep{dejong10}, AEGIS \citep{yoon18}, and H3 \citep{conroy19} have revealed chemically distinct inner and outer halo populations, often with different kinematics and metallicities \citep{yoon18, conroy19}, and in many cases dominated locally by debris from major accretion events like Gaia-Enceladus/Sausage or the Sagittarius stream \citep{fattahi19, malhan25}. This diversity mirrors the stochasticity predicted by simulations, where a few massive progenitors can dominate chemical trends in some halos, while others are assembled from numerous smaller systems \citep{monachesi19, gonza25}. Moreover, chemical abundance patterns, such as [$\alpha$/Fe]-[Fe/H] relations, offer further constraints on progenitor star formation histories beyond what mean metallicity alone can provide \citep{gonza25, grand18}. These results make clear that the metal distribution in stellar halos is not spherically symmetric and that its structure encodes detailed information about past accretion and enrichment.

Observations of the Milky Way add further intrigue. \cite{lian23} measured the integrated stellar metallicity profile of our Galaxy, finding a distinctive $\Lambda$-shaped form: mildly positive inside $\sim$7\,kpc, then steeply negative beyond. Such a profile is not common among Milky Way-mass galaxies in the MaNGA survey or the TNG50 simulation, suggesting that the Milky Way’s enrichment history may be atypical. Meanwhile, chemo-dynamical dissections of halo populations (e.g. Gaia-Sausage/Enceladus) indicate that metallicity is intertwined with kinematics, pointing to distinct accretion epochs and progenitor masses \citep{malhan25, pu25}. Further, the shape of the Metallicity distribution function (MDF) provides a statistical fingerprint of the underlying stellar populations and their distinct formation pathways. Several studies analyzed the metallicity distribution function of the Galactic halo \citep{schorck09, deokkeun13, lamers17, mori24} and M31's stellar halo \citep{ibata01, kalirai06, cohen18}. A single-peaked MDF might suggest a single, dominant formation process, while a multi-peaked distribution points to a composite origin from multiple, chemically distinct populations \citep{carollo07, carollo10}. One can also combine the metallicity distribution with stellar kinematic information to reconstruct the assembly history of the Galactic halo \citep{helmi99, kafle13, belokurov20}. The metallicity-kinematic correlation from observations challenges the assumption of a single homogeneous halo population \citep{deokkeun13}. However, \citep{mori24} pointed out that coupling kinematic information with MDFs to trace the Milky Way’s past mergers can bias the results, making it seem like there were more merger events with smaller progenitor galaxies than there really were. \citep{deokkeun13} 

Despite these advances, most existing studies compress the halo’s rich spatial-chemical structure into one-dimensional radial profiles, losing potentially crucial information about anisotropies and directional substructure. Yet both simulations and observations indicate that metallicity can vary significantly across the sky at fixed radius owing to tidal streams, shells, and asymmetric debris from accretion events. This angular variation has rarely been quantified systematically, and the degree to which metal-rich and metal-poor stars occupy distinct regions of the sky remains poorly understood. Furthermore, while radial metallicity gradients correlate with accretion histories, it is unclear how much additional information about halo assembly is contained in the angular-metallicity correlation.

We aim to address this gap by proposing and applying an information-theoretic framework to quantify the coupling between angular position and metallicity in Milky Way-mass stellar halos from the Aquarius simulations. By treating angular position and metallicity as discrete random variables and computing their mutual information, we can measure the statistical dependence between a star’s sky location and its metallicity class. Unlike traditional profile-based analyses, mutual information captures both global anisotropies and localized chemical substructure in a single, parameter-free measure. Its radial dependence can reveal where spatial-chemical coupling is strongest potentially pinpointing the debris of major mergers, the influence of surviving satellites, or the fossil record of early accretion. Comparing these correlations in the presence and absence of satellites, and against randomized controls, offers a powerful new way to disentangle the roles of in-situ formation, satellite accretion, and dynamical mixing in shaping the present-day halo.

In doing so, we aim to answer fundamental questions: Do metal-rich and metal-poor halo stars cluster differently on the sky at fixed radius? Can this angular-metallicity relationship distinguish halos assembled via distinct accretion histories? And how do these spatial-chemical patterns augment our understanding of halo formation beyond traditional metallicity gradients? By casting the metal distribution problem in terms of spatial entropy and mutual information, our study complements and extends conventional approaches, enriching our interpretative toolkit for upcoming observational surveys.

The structure of this paper is as follows. Section~\ref{sec:data} introduces the simulated stellar halo catalogues and summarizes their key properties. Section~\ref{sec:method} outlines the information-theoretic framework and the quantitative measures used to analyse spatial-chemical correlations. The main results are presented and discussed in Section~\ref{sec:results}, and Section~\ref{sec:conclusion} concludes with a summary of our findings and their broader implications for stellar halo formation.

%%%%%%%%%%%%%%%%%%%%
\section{Data}
\label{sec:data}

Our analysis is based on the mock stellar halo catalogues constructed by \cite{lowing15} from the Aquarius cosmological simulations of Milky Way-mass dark matter halos \citep{springel08a, springel08b, navarro10}. The Aquarius project follows the high-resolution evolution of six dark matter halos (Aq-A to Aq-F) with present-day virial masses of order $10^{12}\,M_{\odot}$. These simulations adopt the standard $\Lambda$CDM cosmology with parameters $\Omega_m = 0.25$, $\Omega_\Lambda = 0.75$, $\sigma_8 = 0.9$, $n_s = 1$, and $h = 0.73$, the same as those used in the Millennium simulation and consistent with the WMAP 1-year and 2dF Galaxy Redshift Survey constraints \citep{spergel03, colless01}. Each halo is simulated at multiple resolution levels. In this work, we use the ``level-2'' runs, which provide an optimal balance between numerical resolution and statistical completeness.

The stellar content of the Aquarius halos is modeled using the particle-tagging scheme introduced by \cite{cooper10}. In this approach, the most tightly bound $1\%$ of dark matter particles in each halo are ``tagged'' to trace the phase-space evolution of stellar populations predicted by the semi-analytic galaxy formation model GALFORM \citep{cole94, cole00, bower06}. Each tagged particle carries information about the age, metallicity, and mass of its corresponding stellar population, thereby linking the simulated dark matter dynamics to the stellar assembly history of the host galaxy.

\cite{lowing15} subsequently converted these tagged particles into fully realized stellar halos by populating each tag with individual stars according to a stellar population synthesis model. We use these mock stellar catalogues for five of the six Aquarius halos (Aq-A through Aq-E). The sixth (Aq-F) was excluded because its recent major mergers at $z\sim 0.6$ likely preclude the presence of a stable disk galaxy at $z=0$. In this scheme, each halo comprises both a diffuse stellar component and a population of bound satellites, all resolved with detailed positional, kinematic, and chemical information for individual stars.

The bound substructures in these halos are identified using the SUBFIND algorithm \citep{springel01}, which detects gravitationally self-bound subhalos within the larger host. By cross-referencing the subhalo IDs associated with the original tagged particles, we separate stars belonging to bound satellites from those that make up the smooth halo and diffuse debris field. This separation allows us to investigate how satellite removal alters the spatial and chemical coherence of the stellar halo.

The metallicity of each star/stellar population is provided in the catalogues as the total mass fraction of all elements heavier than helium. This definition enables a consistent comparison of chemical enrichment across progenitors and between halos. In the present work, these mock stellar halos serve as a physically motivated testbed for applying information-theoretic measures particularly the mutual information between angular position and metallicity to quantify spatial-chemical correlations and to trace how hierarchical accretion leaves persistent informational signatures in stellar halos.

%%%%%%%%%%%%%%%%%%%%

\section{Method of analysis}
\label{sec:method}

Our analysis focuses on quantifying how the angular distribution of stars in stellar halos is connected to their metallicity, and on understanding how these spatial-metallicity correlations vary with distance from the halo centre. We use five mock stellar halos (Aq-A to Aq-E) from the Aquarius simulation \citep{springel08a, cooper10, lowing15}, each representing a Milky Way-mass galaxy. The simulations provide individual stellar particles tagged with positions, metallicities, and stellar masses. Bound satellites are identified using the SUBFIND algorithm \citep{springel01}, allowing us to carry out the analysis both with and without these satellites to isolate their role in driving the observed trends.

For each halo we first determine the median metallicity ($\log_{10}(Z/Z_\odot)_{med}$) of the entire stellar halo (including satellites). Here $Z_{\odot}=0.0139$ \citep{asplund21} is the solar metallicity. Stars with metallicity greater than this median are labeled ``high metallicity'', while those less than or equal are labeled ``low metallicity''. This binary split produces two complementary subsamples whose spatial and statistical properties can be compared directly. Rather than imposing a fixed metallicity threshold to separate ``low''- and ``high''-metallicity stars, we divide each halo at its own median metallicity. This choice is driven by the statistical properties of the information-theoretic measures used in this work. Both the information entropy and the mutual information are sensitive to Poisson fluctuations in the underlying number counts. A predefined metallicity cut would generally yield highly unequal population sizes that differ not only from halo to halo but also as a function of radius, thereby amplifying radius-dependent Poisson noise in an uncontrolled way. By adopting a median-based split, we ensure that the two metallicity classes contain comparable numbers of stars globally, which keeps their overall noise levels comparable and prevents either population from becoming systematically under-sampled. The cumulative counts of the two groups still vary with radius as expected from the intrinsic chemical structure of each halo. However, the median split avoids the far more severe imbalances that a fixed metallicity threshold would introduce, allowing the resulting entropies and mutual informations to reflect genuine spatial-chemical structure rather than artifacts of sample-size asymmetry. The use of cumulative counts in the information-theoretic measures employed in this work is partly motivated by this consideration, as it stabilizes the probability distributions at large radii and mitigates the disproportionate Poisson noise that would otherwise arise from highly imbalanced metallicity samples.

\subsection{Whole-sky anisotropy of low- and high-metallicity stars}
\label{subsec:aniso}
The spatial anisotropy at a given radius is quantified using the whole-sky anisotropy parameter introduced by \cite{pandey16, pandey22} and described in Section~2.2.1 of \cite{mondal24}. The halo is observed from its centre, and the sky is pixelated into $N_{\mathrm{pix}} = 12\,N_{\mathrm{side}}^2$ equal-area HEALPix cells \citep{gorski05}. For any chosen cumulative radius $r$ (all stars with galactocentric distance $\le r$), we count the number of stars $n_i(r)$ in pixel $i$, weighting each particle by its stellar mass. The total number of stars within $r$ is $N(r) = \sum_{i=1}^{N_{\mathrm{pix}}} n_i(r)$. The information entropy associated with the angular distribution is
\begin{equation}
H_{\mathrm{whole-sky}} (r)= \log N(r) - \frac{1}{N(r)} \sum_{i=1}^{N_{\mathrm{pix}}} n_i(r)\,\log n_i(r),
\end{equation}
which reaches a maximum value $H_{\mathrm{max}} = \log N_{\mathrm{pix}}$ for a perfectly isotropic distribution. The whole-sky anisotropy parameter is then defined as
\begin{equation}
a(r) = 1 - \frac{H_{\mathrm{whole-sky}}(r)}{H_{\mathrm{max}}}.
\end{equation}
We compute $a(r)$ separately for all stars, for the low-metallicity subset, and for the high-metallicity subset. This allows us to examine whether the two metallicity populations differ in their angular clustering as a function of radius, and whether such differences persist after removing the bound satellites.

\subsection{Mutual information between angular position and metallicity} 
We treat angular position and metallicity as two discrete random variables:  
\(X\) is the HEALPix pixel index, \(1 \le i \le N_{\mathrm{pix}}\), and \(Z\) is the metallicity label, ``low'' or ``high'' (\(j=1,2\)).  
For stars within a radius $r$ , the joint, mass-weighted counts \(n_{ij}(r)\) give the probabilities
\[
p(X_i) = \frac{\sum_j n_{ij}(r)}{N(r)}, \quad
p(Z_j) = \frac{\sum_i n_{ij}(r)}{N(r)}, \quad
p(X_i,Z_j) = \frac{n_{ij}(r)}{N(r)}.
\]
The marginal entropies, joint entropy, and mutual information \citep{shannon48, pandey17} evaluated within radius \(r\) are
\begin{align}
H_r(X) &= - \sum_{i=1}^{N_{\mathrm{pix}}} p(X_i) \log p(X_i), \\
H_r(Z) &= - \sum_{j=1}^{2} p(Z_j) \log p(Z_j), \\
H_r(X,Z) &= - \sum_{i=1}^{N_{\mathrm{pix}}} \sum_{j=1}^{2} p(X_i,Z_j) \log p(X_i,Z_j), \\
I_r(X;Z) &= H_r(X) + H_r(Z) - H_r(X,Z).
\end{align}
Here the subscript \(r\) simply indicates that the quantity is calculated for all stars with galactocentric radius \(\le r\).

We compute anisotropy and mutual information profiles using all stars within a given radius \(r\) rather than in independent spherical shells. This approach ensures that the number of particles contributing to each measurement grows with radius, which reduces shot noise and stabilizes entropy and MI estimates. It also avoids the strong fluctuations that narrow shells can produce when dominated by individual substructures, making the radial trends easier to interpret. While cumulative profiles are not statistically independent between neighbouring \(r\) values, the improved stability is essential for detecting subtle correlations, particularly in the low-density outer halo.

To determine whether the measured $I_r(X;Z)$ reflects genuine structure rather than chance alignment, the metallicity labels are randomly shuffled among the stars within \(r\), preserving the total numbers of low- and high-metallicity stars but erasing positional correlation. Repeating these randomizations many times yields a reference distribution of MI values expected in the absence of true correlation, against which the real profiles can be compared.

We repeat all anisotropy and MI measurements for both the original halos (with satellites) and for versions in which bound satellites are removed. Comparing these two cases identifies the role of satellites in driving spatial-metallicity correlations. Since satellites are known to dominate the anisotropy signal beyond $\sim 60\,h^{-1}\,\mathrm{kpc}$ \citep{mondal24}, this step is crucial for separating their contribution from that of diffuse substructures and the overall halo shape.

\section{Results}
\label{sec:results}

%%%%%%%%%%%%%%%%%%%%%%%

\subsection{Metallicity distribution functions (MDF) of Aquarius stellar halos} %requires a synthesis paragraph

\begin{figure}
    \centering
    \includegraphics[width=\linewidth]{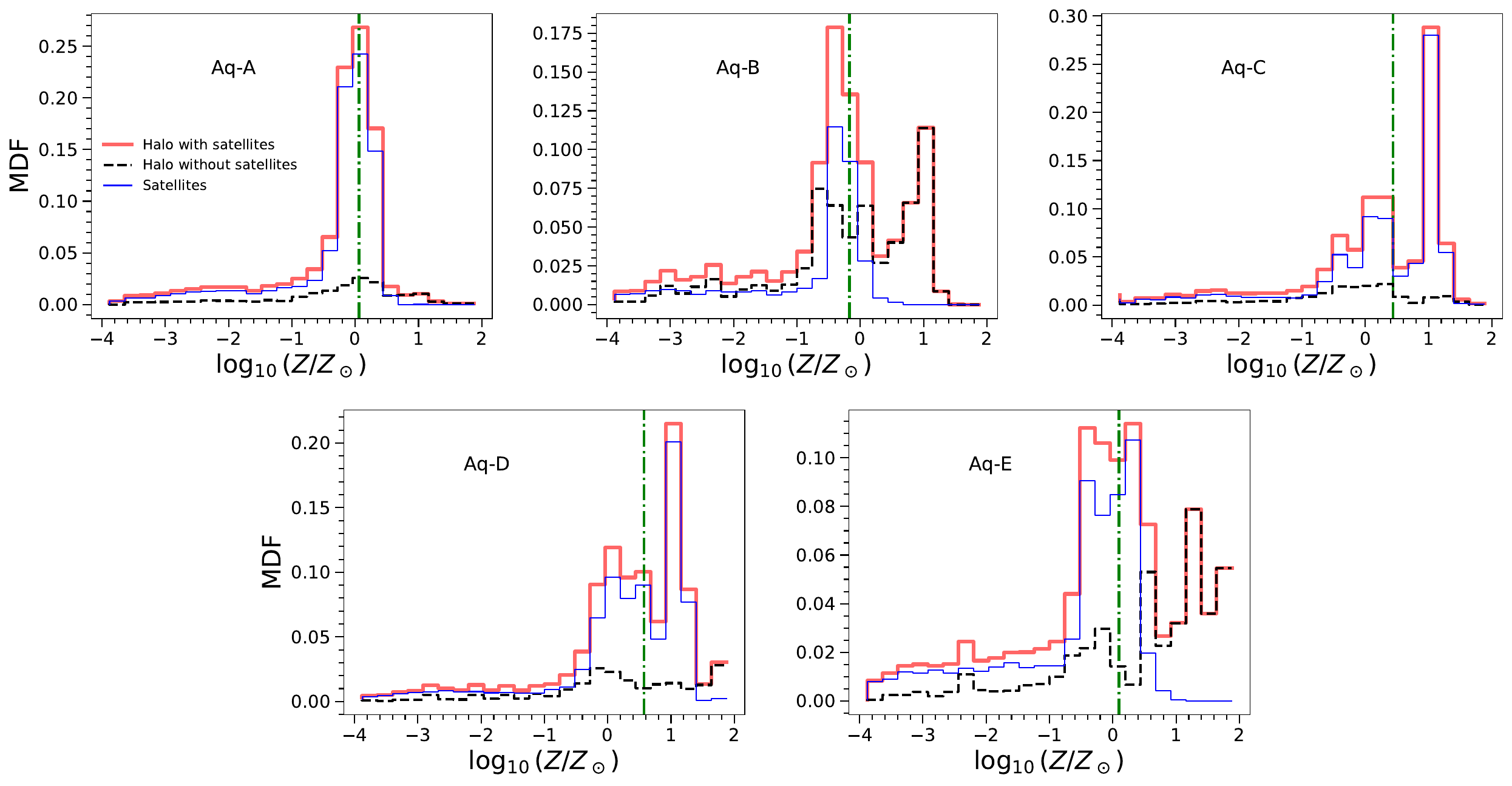}
\caption{This shows the metallicity distribution functions (MDFs) for the five Aquarius stellar halos (Aq-A to Aq-E). The red curves show the MDF of the total stellar halo, the black curves correspond to the smooth halo after removing bound satellites, and the blue curves represent the contribution from satellites alone. In each panel, the vertical dot-dashed green line marks the median metallicity of the entire stellar halo (including satellites) used to divide stars into low- and high-metallicity populations in our analysis. The diversity in MDF shapes ranging from single-peaked to bimodal and satellite-dominated reflects the diverse chemical and dynamical assembly histories of the five halos.}
\label{fig:mdf_aquarius}
\end{figure}

\autoref{fig:mdf_aquarius} presents the metallicity distribution functions (MDFs) for the five Milky Way-mass stellar halos from the Aquarius simulations. In each panel, the total MDF of the stellar halo (including both field and satellite stars) is shown in red, the MDF for the halo after removing bound satellites is shown in black, and the contribution from the satellites alone is plotted in blue. The red distribution thus represents the combined metallicity distribution of both components. This decomposition allows us to isolate the chemical signatures of the accreted and in-situ channels in each system and to trace how these two formation modes together sculpt the final MDF. However, the Aquarius simulations are dark-matter only simulations where stellar halos are built via semi-analytic tagging of dark matter particles based on subhalo histories \citep{cooper10}. All stars in Aquarius are of accreted origin and there is no explicit in-situ component.

The MDF of Aq-A shown in the top left panel of \autoref{fig:mdf_aquarius} is narrow and single-peaked, but importantly the total distribution (red) closely follows the satellites (blue) rather than the non-satellite halo (black). In other words, surviving or recently accreted satellites dominate the overall metallicity budget in this system, imprinting their chemical signature on the global MDF. The apparent unimodality of Aq-A’s MDF therefore does not arise from any smoothly enriched in-situ component, but rather from the cumulative contribution of a few chemically similar, massive progenitors whose early accretion and comparable enrichment histories produced a well-defined, single-peaked metallicity distribution. In this sense, Aq-A represents a halo where the collective imprint of accreted systems rather than internal star formation sets the global metallicity fingerprint.

In contrast, Aq-B in the top middle panel of \autoref{fig:mdf_aquarius} displays a clearly double-peaked MDF, revealing the coexistence of two chemically distinct stellar populations. The lower-metallicity peak likely traces the debris of numerous low-mass satellites accreted over extended timescales, whereas the higher-metallicity peak arises from one or a few massive progenitors that were accreted early and deposited their stars deep within the main halo’s potential well. Interestingly, the high-metallicity wing of the MDF is contributed almost entirely by the smooth halo component, while the satellites populate mainly the lower-metallicity peak. This chemical bifurcation thus reflects the combined imprint of multiple accretion events involving progenitors of very different masses and enrichment histories, a striking example of how hierarchical accretion alone can reproduce the duality often ascribed to both in-situ and accreted formation channels.

The halos Aq-C and Aq-D in top right and bottom left panels of \autoref{fig:mdf_aquarius} also exhibit broad or weakly bimodal MDFs, but with somewhat different proportions of metal-rich and metal-poor components. Their distributions suggest that both underwent complex assembly histories involving the superposition of multiple accretion events spanning a wide range of progenitor masses and enrichment levels. In these systems, the blue curves representing the satellite contributions dominate over the black curves across nearly the entire metallicity range, indicating that the global metallicity distribution is shaped almost entirely by stars originating in accreted systems. At the high-metallicity end, the blue curves almost perfectly overlap with the total red MDFs, showing that even the more metal-rich populations in these halos are contributed predominantly by the disrupted satellites themselves. This dominance of massive, chemically evolved progenitors implies that Aq-C and Aq-D were assembled largely through the accretion of such systems, whose debris collectively defines both the metal-poor and metal-rich ends of their stellar halo metallicity distributions.

Aq-E in the bottom right panel of \autoref{fig:mdf_aquarius} presents a distinct case compared to the other halos. Its MDF is notably broad and exhibits a pronounced skew toward low metallicities, reflecting a chemically diverse but strongly accretion-dominated origin. The blue curve, representing the satellites, dominates the low-metallicity end and extends across a wide range, indicating that numerous low-mass, metal-poor systems have contributed significantly to the halo’s buildup. The total red distribution is therefore largely shaped by the cumulative contribution of these satellites. However, at the high-metallicity end, a small shoulder arises that is entirely due to the smooth halo component (black curve), with no corresponding contribution from surviving satellites. This feature suggests that Aq-E, while overwhelmingly assembled through accretion, also contains a minor fraction of stars originating from one or a few relatively massive progenitors whose debris was deposited early and deep within the halo’s potential well. These centrally deposited remnants, though accreted, produce a modest enhancement in the metal-rich tail, marking the residual imprint of early massive accretion events within an otherwise diffuse, chemically diverse halo.

Together, these MDFs vividly illustrate the diversity of stellar halo assembly pathways within the $\Lambda$CDM framework. A single-peaked MDF can arise through different accretion configurations: in Aq-A, it reflects the dominant imprint of a few chemically similar, massive progenitors accreted early, whereas in other systems a comparable shape might result from the cumulative mixing of many satellites with overlapping enrichment histories. Multi-peaked or skewed distributions, as seen in Aq-B through Aq-E, capture the composite nature of hierarchical buildup, involving contributions from progenitors spanning a broad range of masses and metallicities. The fact that the metal-rich end of the MDF is entirely associated with the smooth halo component in Aq-B and Aq-E underscores that higher-metallicity stars generally trace debris from massive, early-accreted systems deposited deep in the potential well, whereas the low-metallicity wings are shaped by the dispersed remnants of numerous low-mass satellites accreted later. These contrasts collectively highlight how variations in progenitor mass spectrum and accretion timing give rise to the rich diversity of chemical structures observed across stellar halos.

The contrasting MDF shapes across these five halos encapsulate the diversity of stellar halo formation pathways expected in the $\Lambda$CDM paradigm. Within the purely accreted framework of the Aquarius simulations, these differences arise not from distinct in-situ and accreted channels, but from the interplay between progenitor mass, accretion time, and chemical enrichment history. Stars originating in massive progenitors that were accreted early are typically more metal-rich and deposited deep within the host’s potential well, while those from numerous low-mass satellites accreted later are more metal-poor and spatially extended. The varying balance between these components from Aq-A to Aq-E reflects the stochastic nature of hierarchical assembly. Some halos are dominated by the cumulative debris of a few massive systems whereas others are dominated by the dispersed contributions of many smaller ones. Viewed in this light, the MDF serves as a powerful diagnostic of the relative importance of massive versus low-mass accretions in building stellar halos. Its interpretation also provides an essential foundation for connecting chemical substructure, spatial anisotropy, and information-theoretic measures such as mutual information, which together quantify the spatial-chemical complexity imprinted by hierarchical growth.

Overall, the MDFs reveal that the five Aquarius halos span a broad spectrum of assembly histories despite all being formed entirely through accretion. At one extreme, Aq-A is characterized by a relatively simple chemical structure dominated by a few chemically similar progenitors, producing a narrow, unimodal MDF. At the other extreme, Aq-E exhibits a much broader and more metal-poor distribution, indicative of contributions from numerous low-mass systems. Aq-B occupies an intermediate regime where a clear chemical bimodality points to the coexistence of debris from progenitors with markedly different masses and enrichment histories, while Aq-C and Aq-D represent chemically complex halos assembled through the cumulative contributions of multiple significant accretion events. These contrasts demonstrate that the MDF is a powerful probe of the progenitor mass spectrum and chemical enrichment history of a stellar halo. However, the MDF alone contains no information about how stars of different metallicities are distributed across the sky. Distinct assembly histories may therefore produce similar metallicity distributions while leaving very different spatial imprints. To explore this additional dimension of halo structure, we next examine the spatial anisotropy of the low- and high-metallicity populations and investigate how chemical diversity is linked to the angular distribution of stars.

%%%%%%%%%%%%%%%%%%%%%%%

\subsection{Spatial anisotropy of high- and low-metallicity populations} 

\begin{figure}
    \centering
    \includegraphics[width=\linewidth]{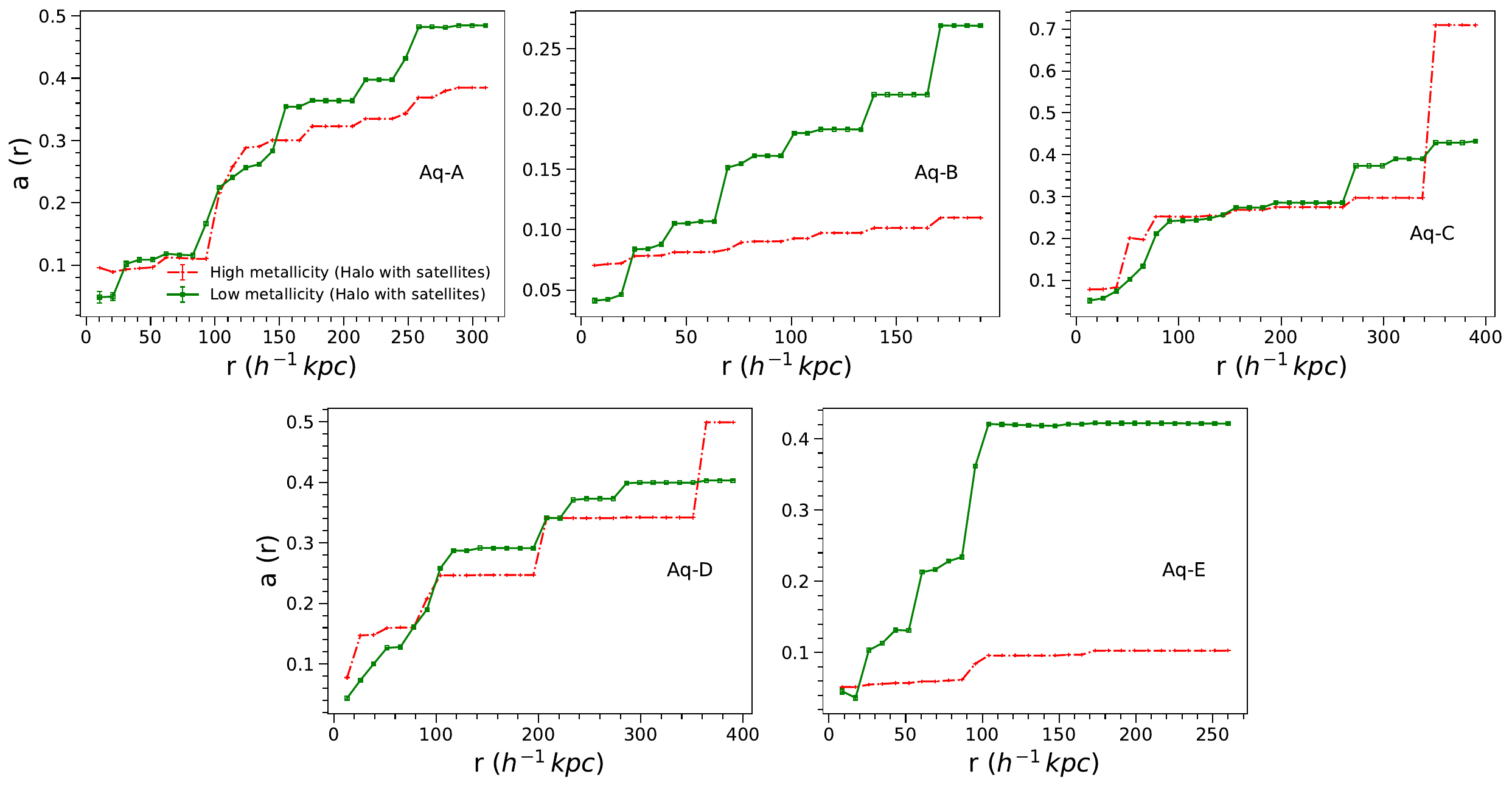}
    \caption{This shows the spatial anisotropy parameter as a function of galactocentric distance for high- and low-metallicity stellar populations in the five Aquarius halos with bound satellites. Each panel corresponds to one halo (Aq-A to Aq-E) including satellites. The low-metallicity stars generally exhibit stronger anisotropy, reflecting their origin in spatially coherent accreted substructures, while high-metallicity stars tend to be more isotropically distributed. The 1$\sigma$ error bars shown here are obtained from 10 jackknife samples drawn from the respective datasets.}
    \label{fig:anisotropy_all}
\end{figure}

We quantified the global anisotropy of the high and low metallicity stellar populations as a function of galactocentric radius using the Shannon-entropy based measure described in \autoref{subsec:aniso}, computed cumulatively out to each radius. The results for the five Aquarius stellar halos (Aq-A through Aq-E, including their satellites) are shown in \autoref{fig:anisotropy_all}.

The anisotropy profiles reveal a rich variety of spatial behaviours, reflecting the underlying diversity of the halos’ assembly histories. In the top left panel of \autoref{fig:anisotropy_all}, both metallicity populations in Aq-A display a comparable level of anisotropy at small radii, but beyond $\sim$100\,kpc the low-metallicity stars exhibit a pronounced rise in anisotropy compared to the high-metallicity component. This pattern suggests that the outer halo of Aq-A is dominated by spatially coherent, metal-poor substructures consistent with our MDF analysis, which showed that Aq-A’s global metallicity distribution is heavily shaped by its satellite population. The higher anisotropy of the metal-poor stars is therefore a natural outcome of their association with disrupted satellites and streams distributed anisotropically across the sky.

The top middle panel of \autoref{fig:anisotropy_all} shows that in Aq-B the anisotropy profiles of the two populations in Aq-B diverge even more clearly. The low-metallicity stars show higher anisotropy at nearly all radii, peaking around 150\,kpc, while the high-metallicity component remains relatively isotropic. This behaviour mirrors the chemical bimodality seen in the MDF of Aq-B and reflects contributions from progenitors of differing mass and accretion histories. The high-metallicity peak likely traces debris from one or a few massive progenitors that were accreted early and deposited stars deep within the potential well, leading to a more spatially mixed and isotropic distribution. In contrast, the low-metallicity stars originate predominantly from numerous low-mass satellites accreted later, whose debris remains less phase-mixed and thus more anisotropically distributed. Together, these results indicate that Aq-B’s chemical bimodality maps directly onto a spatial dichotomy between metal-rich and metal-poor components, each preserving the dynamical memory of its accretion pathway.

The halos Aq-C and Aq-D (top right and bottom left panels of \autoref{fig:anisotropy_all}) show similar qualitative trends: in both cases, the low-metallicity stars consistently exhibit stronger anisotropy than the high-metallicity stars, particularly in the outer regions (150-350\,kpc). This behaviour indicates that these halos are dominated by accreted material that has not yet been fully phase-mixed, giving rise to persistent angular substructure at large radii. The connection with their MDFs where the satellite contribution dominates across nearly the entire metallicity range further supports the view that their spatial anisotropy stems from the complex superposition of debris from multiple disrupted progenitors. The relatively smoother and more centrally concentrated metal-rich component likely originates from a few massive satellites accreted early, whose stars are more spatially mixed within the host potential. Thus, the anisotropy patterns of Aq-C and Aq-D trace the dynamical memory of hierarchical accretion rather than any contribution from in-situ formation.

In the bottom right panel of \autoref{fig:anisotropy_all}, Aq-E exhibits the most extreme contrast between the two populations. Its low-metallicity stars display high anisotropy throughout, while the high-metallicity component remains comparatively isotropic and weak at all radii. This is consistent with the broad, satellite-dominated MDF of Aq-E, which indicated a halo built almost entirely through the accretion of low-mass progenitors. The strong and sustained anisotropy of the low-metallicity stars in Aq-E thus provides a spatial counterpart to its chemically diverse and predominantly accreted nature.

Taken together, the anisotropy profiles reveal that the chemical diversity encoded in the MDFs is accompanied by a corresponding diversity in the spatial organization of halo stars. While the MDFs identify the relative contributions of progenitors with different enrichment histories, the anisotropy measurements show how the debris from those progenitors is distributed across the sky. A remarkably consistent picture emerges across all five halos: the low-metallicity populations, which predominantly trace stars contributed by numerous low-mass satellites, retain a stronger angular anisotropy and therefore preserve a clearer memory of the accretion process. In contrast, the high-metallicity populations are generally more isotropic, reflecting their association with a smaller number of massive progenitors whose debris has undergone greater dynamical mixing. The systematic correspondence between metallicity and anisotropy demonstrates that the stellar halos retain a coupled chemical and spatial record of their assembly history. However, anisotropy alone cannot determine whether the observed spatial structure is directly linked to metallicity or simply reflects the overall distribution of stars. To quantify this connection explicitly, we next examine the mutual information between angular position and metallicity, which provides a direct measure of the degree to which spatial and chemical structures are statistically coupled within stellar halos.

%%%%%%%%%%%%%%%%%%%%%%%

\subsection{Residual spatial anisotropy in the smooth stellar halos (after satellite removal)} 

We recomputed the median metallicity for each halo after removing all bound satellites from them. This ensures that the division into low- and high-metallicity stars reflects the intrinsic chemical distribution of the diffuse halo itself, rather than being biased by the metallicity contributions of surviving satellites. All subsequent anisotropy and mutual-information measurements for the smooth halos use this satellite-cleaned median as the basis for the metallicity split.

\begin{figure*}
    \centering
    \includegraphics[width=\linewidth]{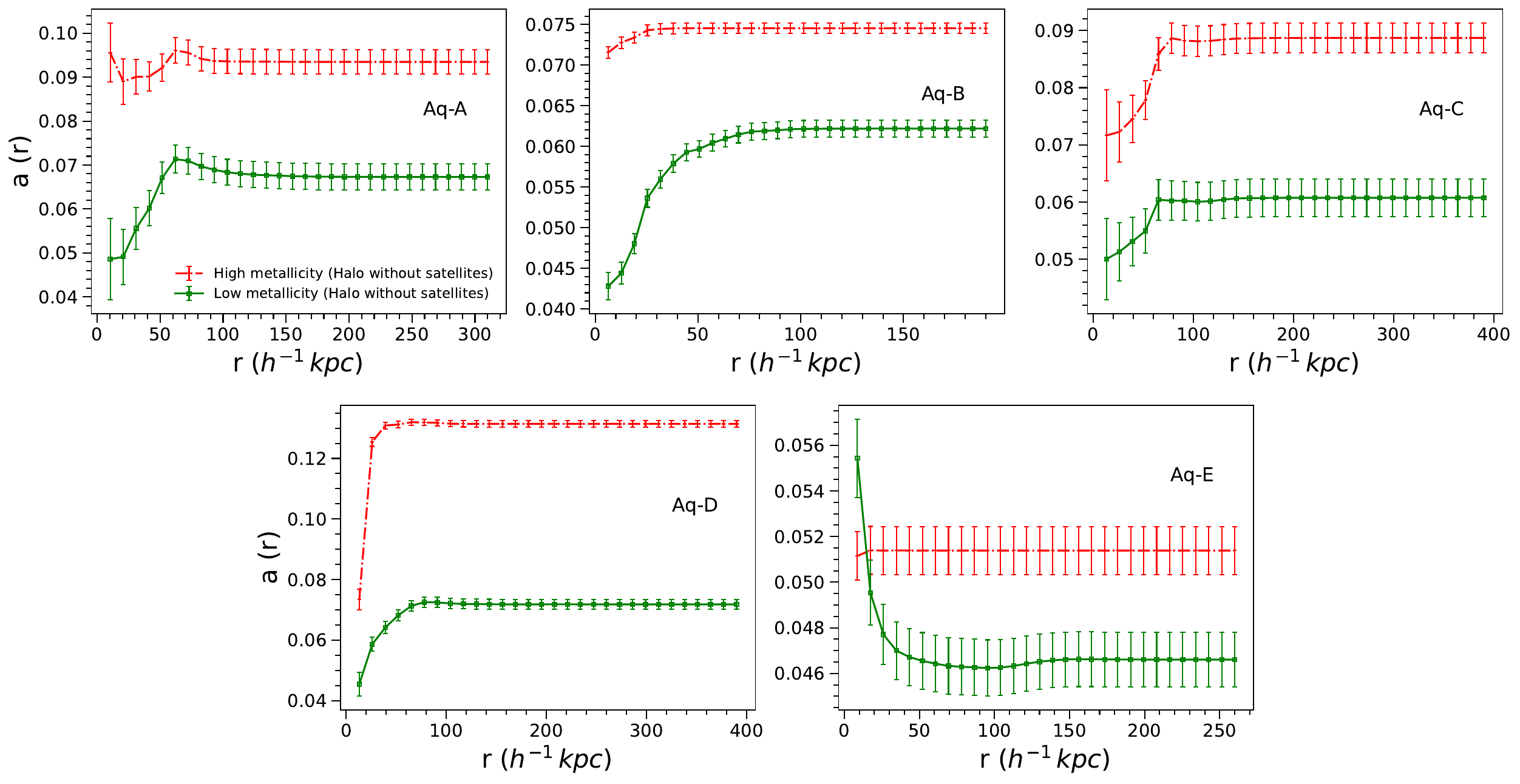}
\caption{This shows the spatial anisotropy parameter as a function of galactocentric distance for high- and low-metallicity stellar populations in the five Aquarius halos after removing bound satellites. Each panel corresponds to one halo (Aq-A to Aq-E). The overall anisotropy decreases significantly after satellite removal, but a subtle trend remains: high-metallicity stars are systematically more anisotropic than their low-metallicity counterparts, reflecting the less isotropic spatial distribution of stars originating from a few massive, early-accreted progenitors. The 1$\sigma$ error bars are computed using 10 jackknife resamples generated from each dataset. }
\label{fig:combine_s0_anis}
\end{figure*}

\autoref{fig:combine_s0_anis} shows the radial variation of spatial anisotropy for the high- and low-metallicity stars in the five Aquarius stellar halos after removing all bound satellites. The removal of satellites markedly reduces the overall anisotropy amplitude across all halos, confirming that much of the angular inhomogeneity seen previously (\autoref{fig:anisotropy_all}) originated from surviving satellites and coherent tidal features. However, a small but systematic differences between the metallicity subsamples persist, revealing residual signatures of the halos’ internal formation and mixing histories.

In all halos, the anisotropy of both metallicity populations remains low indicating that the smooth stellar halo is generally well mixed but a consistent pattern emerges: the high-metallicity stars tend to be slightly more anisotropic than the low-metallicity ones. This trend is evident in all halos from Aq-A to Aq-D, where the high-metallicity population maintains higher anisotropy across nearly all radii. In Aq-E this behavior is confined to the region $>25$\,kpc. The weaker anisotropy of the low-metallicity component suggests that these stars, predominantly originating from numerous low-mass progenitors accreted early, have undergone more efficient phase mixing.

The residual anisotropy of the high-metallicity stars points to their association with the debris of massive early-accreted progenitors whose stars were deposited more centrally and have not been fully isotropized. The persistence of the anisotropy differences between high and low metallicity populations despite the absence of satellites implies that some spatially coherent features are fossil imprints of early accretion events rather than bound substructures.

Overall, the comparison between \autoref{fig:anisotropy_all} and \autoref{fig:combine_s0_anis} highlights a subtle but revealing shift: once satellites are excluded, the spatial anisotropy of the smooth halo becomes low but nonzero, with a systematic enhancement for the metal-rich stars. This reversal relative to the full-halo case underscores the dual nature of stellar halo formation where the low-metallicity stars trace the well-mixed remnants of ancient accretion events, the more metal-rich populations retain the structural memory of their comparatively recent and centrally concentrated origins.

Most importantly, these results show that the removal of bound satellites does not completely erase the spatial signatures of hierarchical assembly. Instead, it reveals a low-level but remarkably consistent anisotropy pattern embedded within the smooth stellar halo itself. Across all five systems, the residual anisotropy is weak, indicating substantial dynamical mixing, yet the high-metallicity populations remain systematically more anisotropic than the low-metallicity ones. This suggests that the smooth halo is not a featureless background, but rather retains the fossilized imprint of a few massive progenitors whose debris was deposited deep within the host potential and has not been fully isotropized. In contrast, the low-metallicity stars appear to trace the cumulative debris of numerous lower-mass accretion events whose remnants have been more thoroughly mixed over cosmic time. The persistence of these metallicity-dependent anisotropy differences demonstrates that information about progenitor mass and accretion history survives even after the most conspicuous substructures have been removed. This naturally raises the question of whether the residual spatial structure is statistically linked to the chemical properties of the stars themselves, a possibility that can be tested directly through the mutual information analysis presented in the following sections.

%%%%%%%%%%%%%%%%%%%%%%%

\subsection{Mutual information between angular position and metallicity}  

\begin{figure*}
    \centering
    \includegraphics[width=\linewidth]{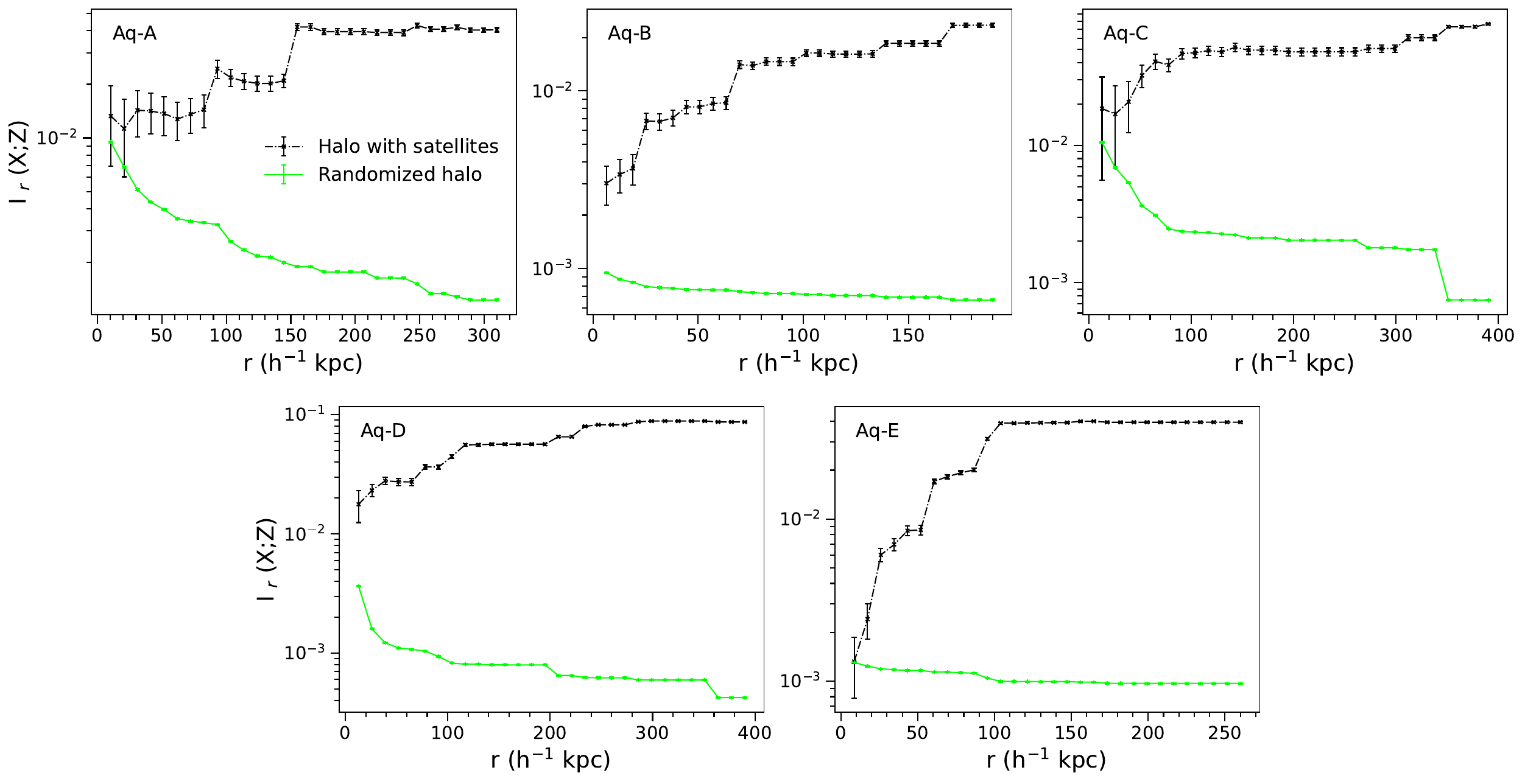}
    \caption{This shows the mutual information $I_{r}(X;Z)$ between angular position ($X$) and metallicity ($Z$) as a function of galactocentric distance for the five Aquarius stellar halos (Aq-A to Aq-E), including satellites. The dot-dashed black lines show the actual halos, while the solid green lines correspond to randomized controls with shuffled metallicity labels. The steady increase and divergence from the randomized curves demonstrate that spatial-chemical correlations strengthen with radius, tracing the growing dominance of anisotropic, chemically distinct accreted substructures. The outer plateau marks the transition to a smoother, dynamically mixed halo component. The 1$\sigma$ errorbars are derived from 10 jackknife samples taken from the underlying datasets.}
    \label{fig:mutual_info}
\end{figure*}

\autoref{fig:mutual_info} presents the radial variation of the mutual information, $I_{r}(X;Z)$, between the angular position ($X$) and metallicity class ($Z$) for the five Aquarius stellar halos, each including their satellite systems. The mutual information quantifies how much information about a star’s metallicity is contained in its angular position, and vice versa. In this framework, $I_{r}(X;Z)$ serves as a global, model-independent measure of spatial-chemical coupling indicating whether metal-rich and metal-poor stars tend to occupy distinct regions of the halo or are spatially intermixed. To establish whether the observed values of $I_{r}(X;Z)$ reflect genuine structure or random fluctuations, we generated randomized control halos in which the metallicity labels of the stars were shuffled while preserving their overall proportions within each radius. The corresponding results are shown alongside the original halos for direct comparison.

In all five halos, the measured $I_{r}(X;Z)$ rises systematically with increasing galactocentric radius, while the randomized controls decline outward. It should be noted that the finite non-zero mutual information in the randomized control originates exclusively from Poisson noise, which diminishes as the number counts increase with radius. The growing divergence between the two curves with radius indicates that spatial-chemical correlations become stronger in the outer halo, where anisotropic substructures such as tidal debris, shells, and streams dominate. Beyond a certain radius, however, this difference tends to plateau, suggesting that the outermost regions contribute little additional structure. This asymptotic behavior likely marks the transition from the structured inner halo, shaped by the remnants of a few massive progenitors, to the diffuse outskirts composed of the debris of many small accretions.

\begin{figure*}
    \centering
    \includegraphics[width=\linewidth]{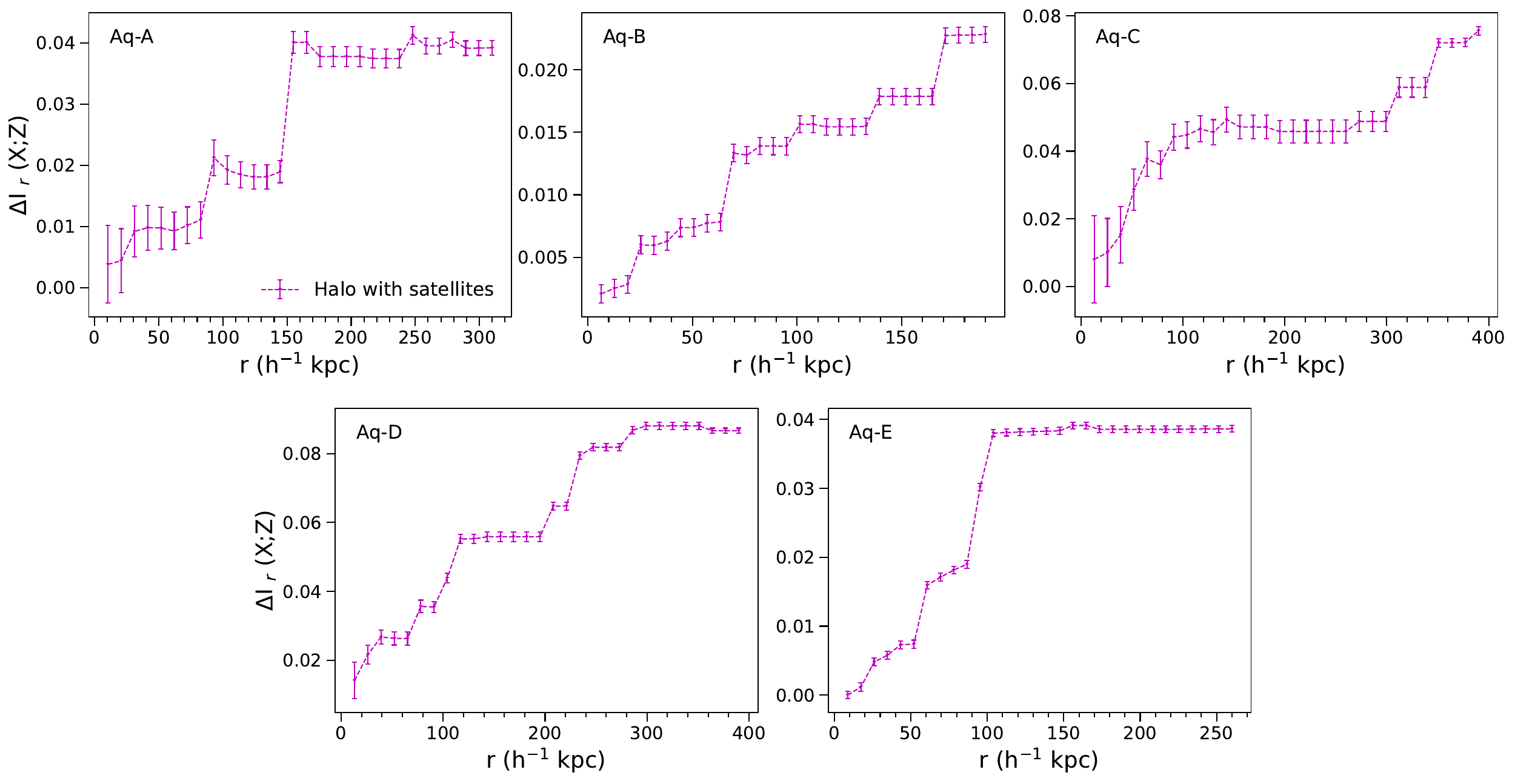}
    \caption{This shows the radial profiles of the difference $\Delta I_{r}(X;Z) = I_{r}^{\mathrm{data}} - I_{r}^{\mathrm{rand}}$ for the five Aquarius stellar halos. This difference isolates the genuine spatial-chemical coupling beyond random expectation. The increasing $\Delta I_{r}(X;Z)$ with galactocentric distance indicates that the outer halos are chemically and spatially more correlated, retaining signatures of their hierarchical assembly.}
    \label{fig:mutual_info_diff}
\end{figure*}

The complementary view is provided in \autoref{fig:mutual_info_diff}, which shows the radial profiles of the difference $\Delta I_{r}(X;Z) = I_{r}^{\mathrm{data}} - I_{r}^{\mathrm{rand}}$ for each halo. This quantity isolates the excess mutual information relative to the randomized baseline, directly highlighting where genuine spatial-chemical coupling exceeds statistical noise. In all cases, $\Delta I_{r}(X;Z)$ increases outward from the galactic center, reflecting the growing dominance of accreted, chemically distinct structures at larger radii. The amplitude and shape of this difference curve, however, vary among halos, revealing key insights into their individual assembly histories.

In Aq-A (top left panel of \autoref{fig:mutual_info_diff}), $\Delta I_{r}(X;Z)$ rises gradually and saturates beyond $\sim150$\,kpc, consistent with a halo assembled primarily from one or two massive progenitors whose debris dominates the outer regions. This behavior aligns with the strong anisotropy of the low-metallicity stars seen earlier (\autoref{fig:anisotropy_all}) and with the MDF dominated by satellite contributions (\autoref{fig:mdf_aquarius}), implying that the outer halo’s spatial coherence and chemical differentiation are primarily driven by the debris of a few large accreted systems. In contrast, Aq-B (top middle panel of \autoref{fig:mutual_info_diff}) exhibits a steeper rise that flattens earlier ($\sim 60$\,kpc), reflecting its chemically bimodal MDF, an imprint of two dominant progenitor populations with distinct enrichment histories and orbital properties. Aq-C and Aq-D (top right and bottom left panels of \autoref{fig:mutual_info_diff}) show higher overall amplitudes of $\Delta I_{r}(X;Z)$, in line with their strongly satellite-dominated MDFs and high spatial anisotropies. Both halos retain pronounced angular-metallicity coupling across large radii, indicating that multiple accretion events with diverse metallicities have deposited overlapping, chemically inhomogeneous debris. Aq-E (bottom right panel of \autoref{fig:mutual_info_diff}), on the other hand, exhibits lower $\Delta I_{r}(X;Z)$ values that plateau at $\sim 100$\,kpc, suggesting that despite being built from numerous low-mass progenitors, its stellar halo has achieved a relatively smooth spatial-chemical configuration through extensive dynamical mixing of its accreted debris.

Together, the mutual information and its excess over the randomized baseline provide a powerful, quantitative lens through which to interpret the hierarchical assembly of stellar halos. The increase of $\Delta I_{r}(X;Z)$ toward larger radii signals the growing contribution of diverse accreted substructures to the cumulative halo, whereas the subsequent plateau shows that the dominant sources of spatial-chemical complexity have been fully accumulated. These trends complement the patterns seen in the MDFs and anisotropy analyses. Whereas the MDFs reveal the chemical diversity of the halo and the anisotropy profiles describe its geometric asymmetry, the mutual information quantifies their interplay showing where and to what extent chemical composition and spatial structure remain statistically entangled. The systematic increase of $I_{r}(X;Z)$ and $\Delta I_{r}(X;Z)$ with radius thus reflects the enduring imprint of hierarchical growth: the outer halo retains the memory of its assembly through the angular and chemical coherence of its accreted stellar populations.

Viewed collectively, the mutual information measurements reveal a level of spatial-chemical organization that is not immediately apparent from either the MDFs or the anisotropy profiles alone. While the MDFs characterize the chemical diversity of the stellar halos and the anisotropy measurements quantify their spatial irregularity, $I_r(X;Z)$ directly measures whether these two properties are physically linked. The systematic excess of mutual information over the randomized expectation demonstrates that metallicity and angular position are not independent quantities in any of the Aquarius halos. Instead, stars with different enrichment histories occupy distinct regions of the halo, preserving a measurable imprint of the accretion events that brought them into the system. The diversity in the shapes and amplitudes of the $\Delta I_r(X;Z)$ profiles further shows that halos with broadly similar levels of chemical complexity can exhibit markedly different degrees of spatial-chemical coherence, reflecting differences in the masses, orbital properties, and accretion epochs of their progenitors. Mutual information therefore provides a genuinely complementary diagnostic of halo assembly, transforming the qualitative notion of chemically distinct substructures into a quantitative measure of how strongly chemical enrichment and spatial distribution remain coupled. An important question, however, is whether this coupling is driven primarily by surviving satellites and recently accreted debris, or whether it persists within the diffuse halo itself. To address this issue, we next examine the mutual information signal after removing all bound satellites.

\subsection{Mutual information in the smooth stellar halos (after satellite removal)} 

\begin{figure}
    \centering
    \includegraphics[width=\linewidth]{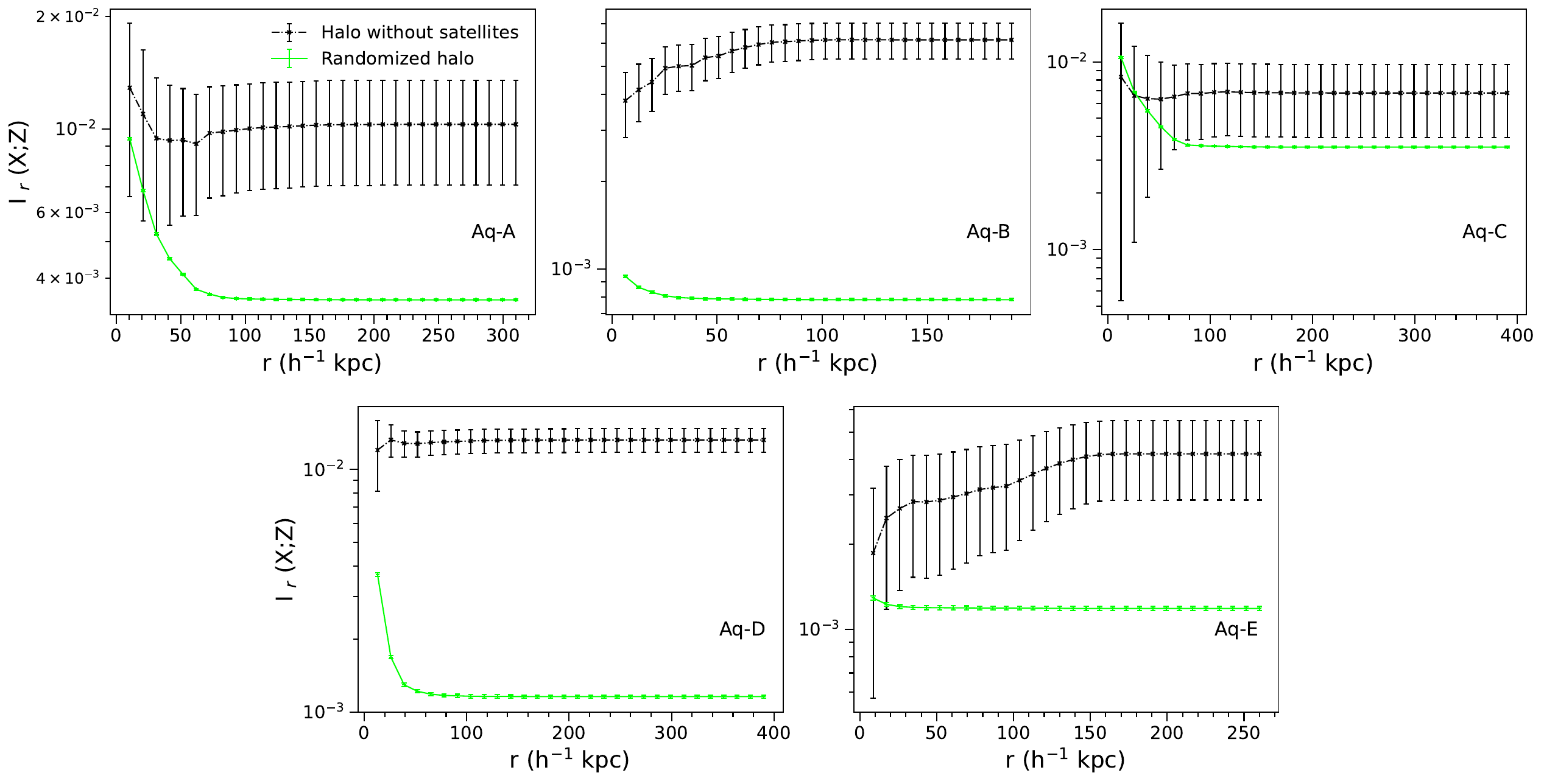}
    \caption{This shows the mutual information $I_r(X;Z)$ between angular position and metallicity as a function of galactocentric distance for the five Aquarius stellar halos after removing bound satellites. Each panel corresponds to one halo (Aq-A to Aq-E). The dot-dashed lines show the results for the halo-only component, while the solid lines represent the randomized counterparts obtained by shuffling the metallicity labels. The overall mutual information decreases significantly after satellite removal, but a residual signal persists in the inner halo, indicating surviving spatial-chemical correlations associated with the debris of massive progenitors. The error bars correspond to 1$\sigma$ uncertainties estimated from 10 jackknife samples of the respective data.}
    \label{fig:mi_s0}
\end{figure}

\begin{figure}
    \centering
    \includegraphics[width=\linewidth]{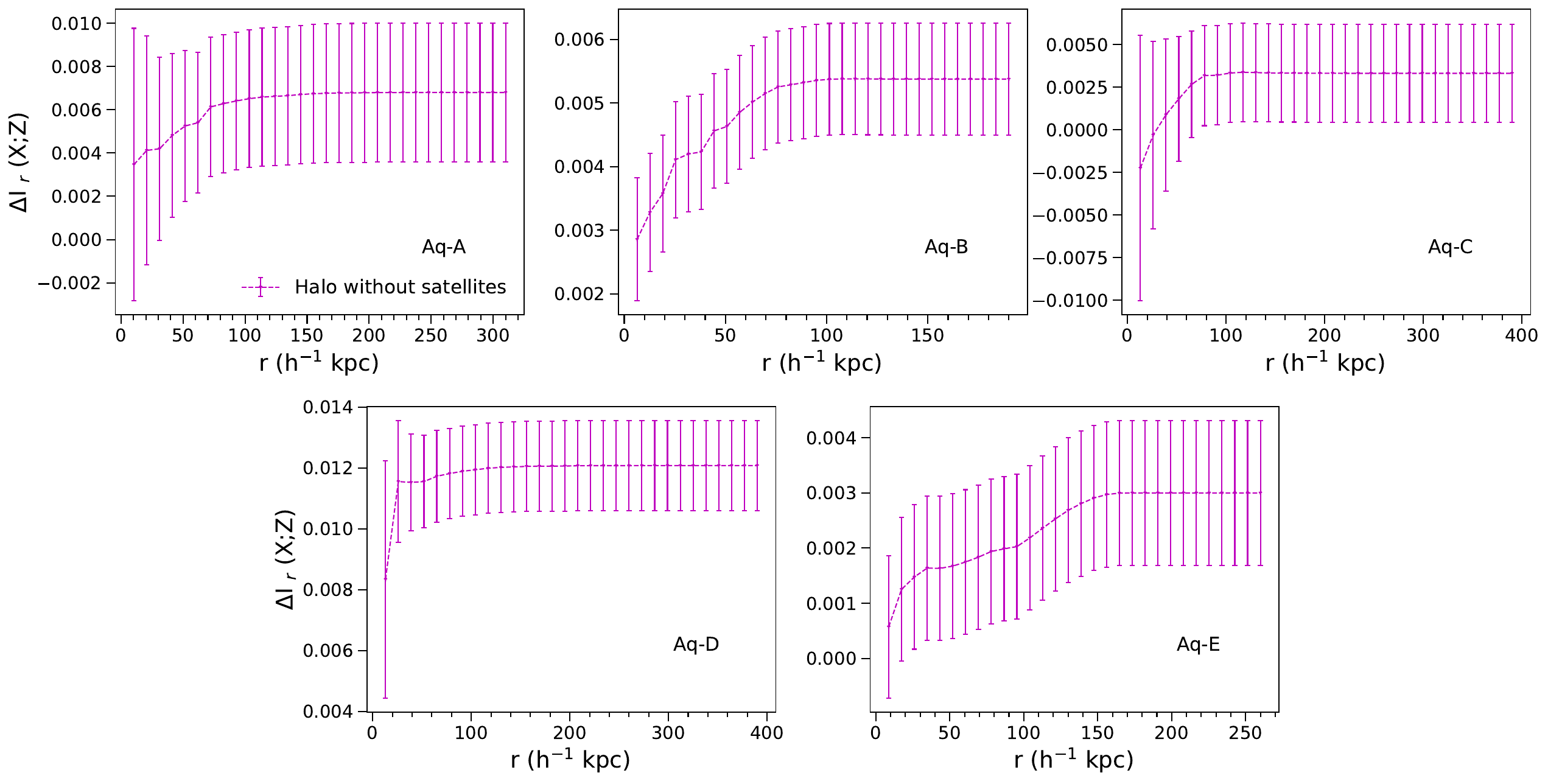}
    \caption{Differences between the measured and randomized mutual information, $\Delta I_r(X;Z)$, for the five Aquarius halos after removing satellites. Each panel corresponds to one halo (Aq-A to Aq-E). The residual $\Delta I_r(X;Z)$ increases with radius in the inner few tens of kiloparsecs and then plateaus, indicating that the remaining spatial-chemical coupling is confined to the central halo regions and largely erased in the outer, well-mixed stellar envelope.}
    \label{fig:mi_diff_s0}
\end{figure}

To assess how the removal of bound satellites affects the spatial-chemical coupling in the stellar halos, we re-computed the mutual information $I_r(X;Z)$ between angular position and metallicity after excluding all stars belonging to self-bound substructures identified by the SUBFIND algorithm. \autoref{fig:mi_s0} shows $I_r(X;Z)$ as a function of galactocentric radius for the five Aquarius halos, with the corresponding randomized controls plotted for comparison. \autoref{fig:mi_diff_s0} presents the difference between the measured and randomized mutual information, $\Delta I_r(X;Z)$, highlighting the radial range where the stellar halos retain genuine spatial-chemical correlations.

After satellite removal, the amplitude of $I_r(X;Z)$ decreases substantially in all systems, demonstrating that a large fraction of the angular-metallicity correlation in the full halos originates from the presence of bound satellites and recently accreted clumps. Nevertheless, a residual signal persists within the inner $\sim 30-50$\,kpc, revealing that even the diffuse stellar halo retains some degree of spatial-chemical coherence. In most halos, $\Delta I_r(X;Z)$ rises with radius in the innermost regions before plateauing beyond a few tens of kiloparsecs, indicating that any remaining correlation is confined to the central parts of the halo and becomes negligible at larger radii.

Aq-A (top left panel of \autoref{fig:mi_diff_s0}) shows a modest rise in $\Delta I_r(X;Z)$ that flattens beyond $\sim$60\,kpc, consistent with its relatively smooth, single-peaked metallicity distribution and weak spatial anisotropy after satellite removal. The residual mutual information likely traces centrally deposited debris from one or two massive progenitors, whose stars dominate both the high-metallicity population and the inner halo structure. In Aq-B (top middle panel of \autoref{fig:mi_diff_s0}), $\Delta I_r(X;Z)$ exhibits a slightly stronger gradient in the inner region, reflecting the chemical bimodality seen in its MDF. The metal-rich debris of the dominant progenitor shapes the inner-halo correlation, while the metal-poor accreted material continues to supply spatial-chemical structure at larger radii, consistent with the cumulative rise and saturation of $\Delta I_r(X;Z)$.

The halos Aq-C and Aq-D (top right and bottom left panels of \autoref{fig:mi_diff_s0}) display relatively higher values of $\Delta I_r(X;Z)$ in their inner regions compared to the other systems, which aligns with their strongly satellite-dominated MDFs and higher anisotropies prior to satellite removal. Their surviving mutual information reflects the overlapping debris of several massive accretions that have not been fully phase-mixed, producing a chemically diverse but spatially correlated inner halo. Beyond $\sim$50$-$80\,kpc the $\Delta I_r(X;Z)$ curves flatten, indicating that most of the spatial-chemical substructure has already been accumulated within the cumulative radius. Aq-E (bottom right panel of \autoref{fig:mi_diff_s0}) shows consistently low mutual information at all radii, reflecting its assembly from numerous low-mass progenitors whose individual contributions are chemically diverse yet spatially diffuse. Because no single accretion event dominates its halo, the cumulative spatial-chemical coherence remains weak throughout.

Comparing these results with the pre-removal trends (\autoref{fig:mutual_info} and \autoref{fig:mutual_info_diff}), it is evident that bound satellites are the dominant contributors to the large-scale spatial-chemical correlations seen in the full halos. Their removal suppresses the overall amplitude of $I_r(X;Z)$ and confines the remaining signal to the inner, more centrally concentrated debris of massive progenitors. The mutual information analysis thus offers a powerful way to disentangle the relative contributions of coherent substructures, massive early mergers, and diffuse, well-mixed components. By quantifying how $I_r(X;Z)$ evolves with radius, we can directly trace the progressive loss of spatial-chemical coherence through dynamical mixing, thereby revealing how the memory of hierarchical assembly fades within the stellar halo.

Overall, these results reveal how the spatial-chemical memory of halo assembly is partitioned between surviving substructures and the diffuse stellar halo. The dramatic reduction in mutual information following satellite removal demonstrates that bound satellites and recently accreted debris are the primary carriers of spatial-chemical coherence in the full halos. Yet the persistence of a residual signal in the inner regions shows that dynamical mixing does not completely erase the imprint of past accretion events. Instead, the smooth halo retains a subtle but measurable record of its formation history through the spatially correlated debris of a few massive progenitors that were accreted early and deposited deep within the host potential. The contrast between the full and satellite-removed halos therefore provides a particularly revealing view of hierarchical growth: recent and surviving accretions dominate the large-scale angular-metallicity correlations, while the weaker residual signal in the diffuse component traces the long-lived fossil remnants of ancient mergers. Mutual information is uniquely sensitive to both regimes, allowing it to distinguish between actively preserved substructure and the more deeply embedded signatures of early assembly. In this sense, the evolution of $I_r(X;Z)$ with radius offers a quantitative measure of how the stellar halo progressively loses, yet never entirely forgets, the spatial-chemical imprint of its hierarchical origin.

\subsection{Connecting metallicity, anisotropy, and spatial-chemical coherence}

The results presented in the preceding sections reveal that the metallicity distribution function (MDF), spatial anisotropy, and mutual information are not independent diagnostics of stellar halo structure, but rather complementary manifestations of the same underlying process: hierarchical assembly through the accretion and disruption of satellite galaxies. Each statistic probes a different aspect of this process. The MDF records the chemical diversity of the progenitor population, the anisotropy quantifies how their debris is distributed across the sky, and the mutual information measures the extent to which these chemical and spatial signatures remain coupled. Collectively, they provide a multidimensional view of halo formation that is substantially richer than any single diagnostic alone.

The five Aquarius halos illustrate this complementarity particularly well. Aq-A, with its narrow, unimodal MDF, relatively modest anisotropy, and gradually increasing mutual information, appears to have been assembled from a small number of chemically similar progenitors whose debris remains spatially coherent at large radii. Aq-B exhibits a markedly different character: its bimodal MDF, strong contrast between the anisotropies of the two metallicity populations, and rapidly rising mutual information all point to the coexistence of progenitors with distinct masses, enrichment histories, and orbital properties. Aq-C and Aq-D represent chemically and dynamically complex systems in which broad MDFs, pronounced anisotropies, and high mutual information amplitudes collectively indicate the superposition of multiple significant accretion events. Aq-E, despite possessing the broadest and most metal-poor MDF, displays comparatively weaker spatial--chemical coherence, suggesting that its numerous progenitors have been more thoroughly dispersed and mixed throughout the halo. Thus, halos that may appear superficially similar when viewed through a single diagnostic can exhibit markedly different assembly signatures when their chemical, spatial, and information-theoretic properties are considered together.

A particularly important insight emerging from this analysis is that chemical complexity and spatial coherence are not necessarily equivalent. A broad MDF indicates the presence of stars with diverse enrichment histories, but it does not reveal whether those stars occupy distinct regions of the halo or are thoroughly mixed. Similarly, a highly anisotropic halo may contain substantial substructure without any clear relationship between that structure and stellar metallicity. Mutual information bridges this gap by directly quantifying the statistical coupling between chemistry and spatial distribution. In doing so, it transforms the qualitative notion of ``chemically distinct substructures'' into a measurable quantity that can be compared across halos and across different stages of their dynamical evolution.

The comparison between the full halos and the satellite-removed halos further clarifies the physical origin of this coupling. The strongest spatial-chemical correlations arise from surviving satellites and recently accreted debris, which preserve both their chemical identity and their spatial coherence. However, the persistence of a residual mutual information signal after satellite removal demonstrates that the smooth stellar halo is not entirely featureless. Instead, it retains a fossil record of early accretion events in the form of weak but measurable spatial-chemical correlations associated with the debris of massive progenitors. The gradual reduction of these correlations from the full halos to the smooth halos provides a direct illustration of how dynamical mixing progressively erases, but does not completely destroy, the memory of hierarchical growth.

Remarkably, these inferences are broadly consistent with the independent reconstruction of the Aquarius stellar halos by \citep{cooper10}, despite the fact that our analysis makes no use of progenitor identities, merger trees, or accretion histories. \citep{cooper10} showed that Aq-C and Aq-D received a dominant fraction ($\sim 60$--$70\%$) of their stellar halo mass from stars stripped from surviving satellites. These are precisely the halos that exhibit the strongest anisotropy signatures and the largest mutual information amplitudes in our analysis, indicating the persistence of spatially coherent, chemically distinct debris. In contrast, Aq-B and Aq-E were found to be assembled primarily through numerous lower-mass accretion events at early times, with surviving satellites contributing less than $10\%$ of the final halo mass. Consistent with this picture, Aq-E displays comparatively weak spatial-chemical coherence despite its broad MDF, while Aq-B retains a strong chemical dichotomy but only moderate large-scale coherence. Aq-A occupies an intermediate regime in both studies, showing evidence for the influence of a relatively significant surviving contributor superposed on an older halo assembled predominantly at high redshift. The agreement between these independent approaches suggests that the present-day metallicity distribution, spatial anisotropy, and angular-metallicity coupling preserve genuine signatures of the underlying assembly history.

Viewed from this perspective, stellar halos can be understood as repositories of information about galaxy assembly. Metallicity traces where stars came from, anisotropy traces how they were deposited, and mutual information quantifies how strongly these two records remain linked. The combination of these diagnostics therefore offers a coherent framework for reconstructing the formation histories of stellar halos and for identifying the surviving signatures of past accretion events. More broadly, our results demonstrate that an information-theoretic approach provides a powerful new lens through which to study the fossil record of galaxy formation, revealing aspects of spatial-chemical organization that remain hidden when chemical and spatial information are analysed separately.

Perhaps the most important result of this work is that mutual information provides information that is not contained in the MDF alone. The MDF quantifies the chemical diversity of a stellar halo, but it cannot determine whether chemically distinct populations remain spatially segregated or have been thoroughly mixed by subsequent dynamical evolution. The mutual information measures precisely this missing ingredient: the degree to which metallicity and position remain statistically coupled. This distinction is particularly evident in Aq-E, whose broad MDF indicates substantial chemical diversity, yet whose relatively weak mutual information signal reveals that much of this diversity has been spatially mixed. Conversely, Aq-C and Aq-D exhibit both strong chemical diversity and strong spatial-chemical coherence, implying that their accreted debris retains a clearer memory of its origin. The agreement between these trends and the independently reconstructed assembly histories of \citep{cooper10} suggests that mutual information is tracing physically meaningful signatures of hierarchical growth rather than merely reflecting statistical properties of the MDF. Mutual information therefore complements rather than replaces traditional MDF analyses, extending chemical archaeology into the spatial domain and providing a direct quantitative measure of how strongly the fossil record of galaxy assembly remains encoded in the present-day stellar halo. By combining information on chemical enrichment and spatial organization, it offers a powerful new route for recovering assembly histories directly from present-day stellar populations, even when detailed progenitor information is unavailable.

\section{Conclusions}
\label{sec:conclusion}

In this work, we have introduced and applied an information-theoretic framework to explore the spatial-chemical structure of stellar halos in Milky Way-mass galaxies using the Aquarius simulations. By combining traditional diagnostics like metallicity distribution functions and spatial anisotropy profiles with the mutual information between angular position and metallicity, we have developed a unified and quantitative description of how chemical enrichment and spatial structure intertwine within the hierarchical assembly of stellar halos.

Our analysis of the MDFs reveals a remarkable diversity across the five Aquarius halos, ranging from single-peaked distributions dominated by one or two massive progenitors (Aq-A) to broad and bimodal profiles indicative of multiple, chemically distinct accretion events (Aq-B, Aq-E). These variations reflect the differing balance between a few dominant mergers and the cumulative contribution of many smaller satellites. The spatial anisotropy analysis further showed that low-metallicity stars originating predominantly from disrupted, low-mass progenitors tend to be more anisotropically distributed when satellites are included, tracing the clumpy substructures left by ongoing or recently accreted systems. After removing bound satellites, the overall anisotropy decreases markedly, yet a subtle inversion emerges: high-metallicity stars become slightly more anisotropic, reflecting the lingering imprint of debris from the most massive, centrally deposited progenitors whose remnants remain only partially mixed. Together, these results demonstrate that the chemical and spatial signatures of hierarchical assembly persist long after dynamical relaxation, encoding a lasting memory of each halo’s accretion history.

The mutual information analysis builds directly on these insights but goes beyond them in a crucial way. While the MDF quantifies chemical diversity and the anisotropy measures geometric irregularity, the mutual information $I_{r}(X;Z)$ captures their interplay by quantifying the degree to which the angular position of stars is statistically coupled with their metallicity. Across all five halos, we find that $I_{r}(X;Z)$ increases with galactocentric radius, diverging systematically from randomized controls that erase spatial-chemical correlations. This behaviour reveals that the outer halo retains significant spatial-metallicity coherence, driven by the anisotropic distribution of accreted, metal-poor substructures. The saturation of $I_{r}(X;Z)$ at large radii marks the point at which most chemically distinct debris has been incorporated into the cumulative sample.

These results resonate with, and provide theoretical context for, observational studies of the Milky Way. The present-day Galactic halo is known to comprise two broadly overlapping components, the inner and outer stellar halo \citep{chiba01}. Observations indicate that stars in the inner halo are kinematically and chemically distinct from the surrounding smooth background \citep{schlaufman11}, and Gaia data reveal that much of this component originates from the debris of a massive progenitor, Gaia-Enceladus, accreted early in the Galaxy’s history \citep{helmi18}. Our findings mirror this structure: the rise of $I_{r}(X;Z)$ in the inner regions reflects the contribution of chemically coherent debris from massive progenitors, while its eventual plateau captures the transition to a dynamically older, more diffuse outer component.

The analysis of mutual information after removing bound satellites further reinforces the interpretive strength of this framework. We find that much of the spatial-chemical coupling in the full halos originates from coherent substructures associated with surviving or recently accreted satellites. Once these are excluded, the mutual information declines sharply and becomes confined to the inner $\sim$30-50\,kpc, tracing debris from a few massive progenitors that were accreted early and only partially phase-mixed. This residual signal encodes the lasting memory of major merger events, while its gradual rise and rapid saturation indicate the limited spatial extent of coherent chemical structure in the smooth halo. Together, the before-after comparison of $I_r(X;Z)$ provides a physically intuitive and observationally consistent picture of how hierarchical assembly imprints and gradually erases spatial-chemical coherence in stellar halos.

More broadly, the mutual information framework offers several key advantages. It provides a model-independent, parameter-free measure of correlation that is sensitive to both linear and non-linear relations between chemical and spatial variables. Unlike traditional profile-based analyses, which compress multidimensional data into one-dimensional averages, the information-theoretic approach naturally incorporates the full statistical complexity of halo substructure. It allows one to assess not only the amplitude but also the scale and extent of chemical inhomogeneities as a function of position. Its cumulative definition ensures that it is robust to sampling noise, making it particularly well-suited for both simulations and real survey data with incomplete sky coverage.

The framework developed here has immediate applications beyond the present analysis. In the context of Galactic archaeology, mutual information provides a powerful quantitative tool to assess the spatial-chemical coherence of accreted substructures identified in large stellar surveys such as Gaia \citep{gaia23}, APOGEE \citep{apogee17}, GALAH \citep{galah15}, and WEAVE \citep{weave24}. By extending this analysis to velocity space one can also explore how spatial-chemical correlations are coupled with kinematic properties. The same formalism can be used to trace the temporal evolution of spatial-chemical coupling in cosmological simulations, quantifying how dynamical mixing and successive merger events progressively erase or preserve information from early accretion epochs. More broadly, this information-theoretic perspective provides a unified statistical framework for comparing simulated and observed halos, offering a rigorous, model-independent means of probing how the fossil record of hierarchical assembly is encoded in the joint distribution of stars across space, chemistry, and dynamics.

An important advantage of the information-theoretic framework developed here is its direct applicability to observational studies of the Galactic stellar halo. The quantities employed in this work require only the angular positions and metallicities of stars, both of which are routinely measured by modern surveys. The rapidly expanding combination of astrometric data from \textit{Gaia} and large spectroscopic programs such as APOGEE, GALAH, WEAVE, 4MOST, and DESI will soon provide unprecedented samples of halo stars with well-characterized spatial and chemical properties. In this context, mutual information offers a particularly attractive diagnostic because it is non-parametric and sensitive to both large-scale anisotropies and localized chemical substructures that may be overlooked by traditional one-dimensional statistics. While a detailed assessment of observational uncertainties, survey selection effects, and sky incompleteness lies beyond the scope of the present work, the strong and systematic trends identified in the Aquarius halos suggest that spatial--chemical correlations should be detectable in sufficiently large stellar samples. The framework introduced here therefore provides a promising new avenue for connecting simulations and observations, enabling future studies to quantify how the fossil record of hierarchical assembly is encoded in the joint spatial and chemical distribution of stars.

An important caveat of the present study is that the Aquarius stellar halo catalogues employed here contain only the accreted component of the stellar halo and do not model stars formed in situ within the central galaxy. Consequently, the spatial and chemical structures analysed in this work arise entirely from the debris of disrupted satellites. Observational \citep{nissen10,bonaca17,belokurov18} and theoretical \citep{zolotov09,cooper15,yu20} studies suggest that the inner regions of stellar halos may contain a substantial population of in situ stars, either formed within the main progenitor itself or displaced from the galactic disc during past merger events. Such stars are expected to be more centrally concentrated and, on average, more metal-rich than the accreted halo population. Their presence could therefore modify the metallicity distribution functions, alter the relative contributions of the high- and low-metallicity populations in the inner halo, and potentially introduce additional spatial-chemical correlations. While the information-theoretic framework developed here remains entirely general, the specific trends reported in this work should be interpreted as characterizing the accreted stellar halo component. Extending this analysis to modern hydrodynamical simulations (Auriga,\citep{monachesi19}) that self-consistently model both accreted and in situ stellar populations will provide an important test of the robustness and broader applicability of mutual information as a tracer of galaxy assembly history.

Finally, the mutual information approach opens a new window into the study of stellar halos, revealing how the interplay between angular structure and chemical composition encodes the fossil record of galaxy formation. By quantifying the shared information between spatial and chemical dimensions, this framework moves beyond descriptive morphology toward a statistically grounded, physically interpretable measure of halo complexity. As upcoming surveys continue to map the Milky Way’s outer regions with unprecedented precision, information-theoretic tools such as mutual information promise to become indispensable in disentangling the intertwined histories of star formation, accretion, and dynamical evolution that shaped our Galaxy.

\section{Acknowledgements}
The authors sincerely thank an anonymous reviewer for insightful comments and suggestions that helped to improve the draft. The authors express their gratitude to Carlos Frenk, Andrew Cooper, and Wenting Wang for providing the original tag files of the five Aquarius halos. BP thanks Wenting Wang and Andrew Cooper for their assistance in understanding the data. AM acknowledges the University Grants Commission (UGC), Government of India, for support through a Junior Research Fellowship. BP gratefully acknowledges financial support from the Anusandhan National Research Foundation (ANRF), Government of India through the project ANRF/ARG/2025/000535/PS. BP also acknowledges the Inter-University Centre for Astronomy and Astrophysics (IUCAA), Pune, for support through the Associateship Programme. 

\section{Data availability}
The mock stellar halo catalogues for the five Aquarius halos (Aq-A to Aq-E) used in this study are publicly accessible at \url{http://galaxy-catalogue.dur.ac.uk:8080/StellarHalo/}. The original particle-tag files underlying these mock catalogues are not yet publicly released. Access to them was kindly provided to the authors by Carlos Frenk, Andrew Cooper, and Wenting Wang. Additional data products generated in this work will be made available by the authors upon reasonable request.

%%%%%%%%%%%%%%%%%%%%%%%%%%%%%%%%%%%%%%%%%%%%%%%%%%%%%%%%%%%%%%%%%%%%%%%%%%%%%%%%%%%%%%%%%%%%%%%%%%%%%%%%%%%%%%%%%%%%%%%%%%%%%%%%%%%%%%%%%%%%%%%%%%%%%%%%%%%%%%%%%%%%%%%%%%%%%%%%%%%%%%%%%%%%%%%%%%%

\bibliography{refs}
\bibliographystyle{JHEP}
\end{document}